\documentclass[letterpaper, 10 pt, conference]{ieeeconf}  % Comment this line out if you need a4paper

\IEEEoverridecommandlockouts                              % This command is only needed if 
\usepackage{graphics} % for pdf, bitmapped graphics files
\usepackage{epsfig} % for postscript graphics files
\usepackage{mathptmx} % assumes new font selection scheme installed
\usepackage{times} % assumes new font selection scheme installed
\usepackage{amsmath} % assumes amsmath package installed
\usepackage{amssymb}  % assumes amsmath package installed
\usepackage{mathtools}

\usepackage{caption}
\usepackage{subcaption}

\usepackage{graphicx,xcolor}% http://ctan.org/pkg/{grahpicx,xcolor}

\usepackage{mcode}

\usepackage{xfrac}

\usepackage{amsfonts}
\usepackage{booktabs}
\usepackage{siunitx}

\newlength\mylen
\usepackage{array} % for "\newcolumntype" macro
\newcolumntype{C}{>{\hfil$}p{\mylen}<{$\hfil}} % centered, in math mode, fixed width

\usepackage{cuted}
\usepackage{hyperref}

\title{\LARGE \bf
Optimal assignment of collaborating agents in multi-body asset-guarding games
}

\author{Emmanuel Sin, Murat Arcak, Andrew Packard, Douglas Philbrick, Peter Seiler% <-this % stops a space
\thanks{E. Sin and A. Packard are with the Department of Mechanical Engineering, University of California, Berkeley, e-mail: \{emansin, apackard\}@berkeley.edu. M. Arcak is with the Department of Electrical Engineering \& Computer Sciences, University of California, Berkeley, email: arcak@berkeley.edu. D. Philbrick is with the U.S. Naval Air Warfare Center-Weapons Division, email: douglas.philbrick@navy.mil. P. Seiler is with the Department of Electrical Engineering \& Computer Science, University of Michigan, email: pseiler@umich.edu.}% <-this % stops a space
}   

\begin{document}

\maketitle
\thispagestyle{empty}
\pagestyle{empty}

\author{Emmanuel Sin, Murat Arcak, Andrew Packard, Peter Seiler% <-this % stops a space
\thanks{E. Sin and A. Packard are with the Department of Mechanical Engineering, University of California, Berkeley, e-mail: \{emansin, apackard\}@berkeley.edu. M. Arcak is with the Department of Electrical Engineering \& Computer Sciences, University of California, Berkeley, email: arcak@berkeley.edu. 
P. Seiler is at the Department of Aerospace Engineering and Mechanics, University of Minnesota, email: seile017@umn.edu.}% <-this % stops a space
}

%%%%%%%%%%%%%%%%%%%%%%%%%%%%%%%%%%%%%%%%%%%%%%%%%%%%%%%%%%%%%%%%%%%%%%%%%%%%%%%%
\begin{abstract}

We study a multi-body asset-guarding game in missile defense where teams of interceptor missiles collaborate to defend a non-manuevering asset against a group of threat missiles. We approach the problem in two steps. We first formulate an assignment problem where we optimally assign subsets of collaborating interceptors to each threat so that all threats are intercepted as far away from the asset as possible. We assume that each interceptor is controlled by a collaborative guidance law derived from linear quadratic dynamic games. Our results include a 6-DOF simulation of a 5-interceptor versus 3-threat missile engagement where each agent is modeled as a missile airframe controlled by an autopilot. Despite the assumption of linear dynamics in our collaborative guidance law and the unmodeled dynamics in the simulation environment (e.g., varying density and gravity), we show that the simulated trajectories match well with those predicted by our approach. Furthermore, we show that a more agile threat, with greater speed and acceleration, can be intercepted by inferior interceptors when they collaborate. We believe the concepts introduced in this paper may be applied in asymmetric missile defense scenarios, including defense against advanced cruise missiles and hypersonic vehicles.
\end{abstract}

%%%%%%%%%%%%%%%%%%%%%%%%%%%%%%%%%%%%%%%%%%%%%%%%%%%%%%%%%%%%%%%%%%%%%%%%%%%%%%%%
\section{INTRODUCTION}

The 2019 Missile Defense Review \cite{MDA} describes an evolving threat environment due to advanced cruise missiles and hypersonic weapons that can travel at exceptional speeds with unpredictable flight paths, challenging existing defensive systems. The escalated difficulty posed by such highly maneuverable threats require novel defensive capabilities. The approach studied in this paper is to employ a salvo defense where teams of interceptors under the influence of a collaborative guidance law engage more agile, evasive threats. When multiple threats must be engaged, the teams of interceptors must be optimally assigned to each threat.

In \cite{Isaacs}, Isaacs introduces the classic target-guarding game involving a pursuer, evader, and defender. The author notes conditions on defender maneuverability relative to that of the pursuer required for successful defense of the evader. Since this seminal work, such ``pursuit-evasion'' games and relevant solution methods have been well studied using the theory of dynamic games \cite{Basar1}, \cite{Engwerda}, \cite{Hespanhagame}.

The target-guarding game has been used in missile defense applications such as the problem of protecting an aircraft from a homing missile. In \cite{Garcia1}, \cite{Garcia2}, \cite{Garcia3}, the authors address ``Target-Attacker-Defender'' (TAD) scenarios where the three agents have planar, nonlinear equations of motion. The target and defender cooperate under a guidance law derived from either an optimal control problem or dynamic game where the objective is to maximize the distance between the target and attacker at the time of intercept. In \cite{Prokopov} the authors consider different cooperation schemes between the defender and target, showing that although two-way cooperation provides the best performance, one-way cooperation schemes (e.g., defender assists target but not vice versa) are still better than when the agents act independently.  \iffalse Similarly, \cite{Ratnoo} also compares various guidance strategies and provides analytic launch envelopes and lateral acceleration ratios for the various guidance scenarios.\fi The authors in \cite{Shaferman1} also consider the TAD scenario with planar, nonlinear simulation models but use linearized kinematic models in an adaptive cooperative defender-target guidance scheme. The derived guidance laws induce evasive maneuvers by the target so that the defender's control effort is minimized -  supporting the idea of using relatively inferior defending missiles against potentially more capable attacker missiles. Linearized planar models are also used in \cite{Shima} and \cite{Perelman} to derive cooperative guidance laws in TAD scenarios using optimal control and differential game formulations, showing that linear models are sufficient for guidance applications. 

Other work in target-guarding games include \cite{Shaferman2}  where cooperating pursuers engage a moving target with the objective to not only minimize miss distance but also impose a certain flight path angle relative to teammates. In \cite{Li}, the authors implement a receding horizon algorithm that employs the linear quadratic game framework for target guarding games. The authors of \cite{Kirchner} present a method that can efficiently solve a certain class of Hamilton-Jacobi equations related to the time-optimal guidance control of multiple pursuers collaborating to capture an evader.

In this paper we consider TAD scenarios but whereas the cited prior work generally addresses scenarios with a single threat and a single interceptor, we consider an arbitrary number of collaborating interceptors defending an asset against an arbitrary number of threats - an ``M-vs-N'' multi-body engagement. While existing work on collaborative guidance has focused on cooperation between interceptor and asset or between two interceptors, deriving specific analytical solutions for those special cases and/or using relative reference frames, we introduce a framework that may be generalized to engagements of arbitrary size. Our approach consists of two steps: (1) deriving collaborative guidance laws and (2) predicting the performance of the guidance laws by solving an assignment problem. We use results from linear quadratic dynamic game theory to develop collaborative guidance laws that scale well with large engagements. 

In our assignment problem we employ combinatorial optimization methods developed for general agent-task assignment problems, such as those described in \cite{Kuhn} and \cite{Pentico}. These methods are also used in missile resource allocation problems for independently guided interceptors in \cite{Matlin} and \cite{Karasakal}. However, we extend the problem to the assignment of teams of \textit{collaborating} interceptors against multiple threats. The problem of assigning collaborating agents to tasks was studied in \cite{Younas} for general agent-task problems but to the best of our knowledge has not been applied to missile defense.

\section{PRELIMINARIES}

We use the notation $I_i$ to refer to the $i^{th}$ interceptor and $T_j$ to refer to the $j^{th}$ threat. The set of all interceptors and the set of all threats are denoted as $\mathbf{I} := \{ I_1, \ldots, I_M \}$ and $\mathbf{T} := \{ T_1, \ldots, T_N \}$, respectively. Let $\mathbb{I}_k$ for $k = 1, \ldots, 2^M-1$ denote each of the possible subsets of $\mathbf{I}$, excluding the empty set. We use the notation $\mathbb{I}_k [i]$  to refer to the $i^{th}$ interceptor of interceptor group $k$. As an example, for three interceptors $I_1,I_2,I_3 \in \mathbf{I}$ we may define the following interceptor groups:
\begin{alignat}{5}
	&\mathbb{I}_1 := \{ I_1 \} &&\mathbb{I}_2 := \{ I_2 \} &&\mathbb{I}_3:= \{ I_3 \} &&\mathbb{I}_4 := \{ I_1, I_2 \} \nonumber \\
	&\mathbb{I}_5 := \{ I_1, I_3 \} \ \ &&\mathbb{I}_6 := \{ I_2, I_3 \} \ \ &&\mathbb{I}_7 := \{ I_1, I_2, I_3 \} \nonumber
\end{alignat}
and note that the $2^{nd}$ interceptor in the $6^{th}$ group is $\mathbb{I}_6 [2] = I_3$.

Throughout the paper we use $\mathbf{0}$ to represent a rectangular zero matrix where a superscript refers to the matrix dimensions. Similarly, we use $\mathbf{1}$ to refer to the identity matrix.

\section{ASSIGNMENT PROBLEM FORMULATION}

The mathematical model for the \textit{collaborating agents assignment problem} (hereinafter referred to as the ``assignment problem,'' is given by equations (\ref{cost}) - (\ref{constraint4}). The problem data that must be computed prior to solving this assignment problem is the \textit{reward matrix}, $r \in \mathbb{R}^{\scriptscriptstyle N \times 2^M-1}$, where $r_{\scriptscriptstyle jk}$ is the reward of assigning interceptor group $\mathbb{I}_k$ to threat $T_j$. This elemental reward $r_{\scriptscriptstyle jk}$ is defined in the subsequent section; we consider its computation as a separate issue from the formulation of the assignment problem.
\begin{align}
  &\max_{ z } \ \ \ \min_{j,k} \ \  r_{\scriptscriptstyle jk} z_{\scriptscriptstyle jk} \label{cost} \\
  &\text{ s.t.} \sum_{k=1}^{2^M-1} \hspace{0.8cm} \ z_{\scriptscriptstyle jk} = 1  \hspace{0.55cm} \forall \ j = 1, \ldots, N \label{constraint1} \\
  &\phantom{\text{s.t. }} \ \sum_{j=1}^{N} \hspace{0.9cm} \ z_{\scriptscriptstyle jk} \leq 1 \hspace{0.55cm} \forall \ k = 1, \ldots, 2^M-1 \label{constraint2} \\
  &\phantom{\text{s.t.}} \sum_{ \{ k \mid I_i \in \mathbb{I}_k \} }  \ \sum_{j=1}^{N} \ z_{\scriptscriptstyle jk} \leq 1 \hspace{0.55cm} \forall \ i  = 1, \ldots, M \label{constraint3} \\
  &\phantom{\text{s.t. }} \hspace{1.55cm} \ z_{\scriptscriptstyle jk} \in \{ 0, 1 \} \label{constraint4}
\end{align}

The assignment matrix $z \in \{ 0, 1 \}^{N \times 2^M-1}$ is the decision variable, where $z_{\scriptscriptstyle jk} = 1$ if interceptor group $\mathbb{I}_k$ is assigned to threat $T_j$ and $0$ otherwise. Note that the objective in equation (\ref{cost}) is defined as the smallest entry of the elementwise product between r and z. Hence, we are maximizing the smallest elemental reward that results from assembling and assigning certain interceptor groups to each of the threats.

As shown, constraint sets (\ref{constraint1}) - (\ref{constraint3}) assume that there are more interceptors than threats (i.e., $M \geq N$). The first set of constraints (\ref{constraint1}) ensure that every threat $T_j$ is assigned an interceptor group while the second set of constraints (\ref{constraint2}) enforce each possible interceptor group $\mathbb{I}_k$, from $k = 1, \ldots, 2^M-1$, to pursue no more than one threat. We use the set of constraints in (\ref{constraint3}) to limit each individual interceptor to participate in no more than one interceptor group. Finally, constraint (\ref{constraint4}) ensures that the elements of the matrix $z$ are binary. Note that by introducing a slack variable and transforming the optimization program into epigraph form, we can formulate the stated assignment problem as a mixed-integer linear program to be solved by off-the-shelf solvers.

When $M \geq N$, it is possible to intercept all threats. In fact, if $M$ is strictly greater than $N$, it may not be necessary to launch all interceptors to maximize the objective. However when $M < N$, there are not enough interceptors to intercept all threats. In this case, we assume that all interceptors must be employed to maximize the objective under limited resources. We therefore modify the assignment problem for the $M < N$ case by changing the equality in equation (\ref{constraint1}) into an ``$\leq$'' inequality, and also changing the inequality in equation (\ref{constraint2}) into an equality. An inequality in the constraint equations means that the summation on the left-hand side can equal either $0$ or $1$ whereas an equality requires it to equal $1$.

\section{ASSIGNMENT PROBLEM DATA}

In this section we define the reward matrix, $r$, for our assignment problem and show how we may compute it for an arbitrarily large $M$-versus-$N$ engagement. We first introduce the \textit{single agent model} - the dynamical model used to predict the motion of individual agents - followed by the \textit{multi-agent model}. Then we describe a collaborative guidance law that is used to produce the elemental reward $r_{jk}$ that quantifies that value of assigning interceptor group $\mathbb{I}_k$ to threat $T_j$.  

\subsection{Dynamical Models for Prediction}

\subsubsection{Single Agent Model}

We first assume a double-integrator, continuous-time dynamical model for each of the agents. The state vectors of the (evading) asset, interceptors, and threats are denoted by $x_{\scriptscriptstyle E}$, $x_{\scriptscriptstyle I_i}$, $ x_{\scriptscriptstyle T_j} \in \mathbb{R}^{n_x}$, respectively, where $n_x = 6$ for a 3-dimensional engagement. Similarly, we denote the input vectors as $u_{\scriptscriptstyle E}$, $u_{\scriptscriptstyle I_i}$, $ u_{\scriptscriptstyle T_j} \in \mathbb{R}^{n_u}$, where $n_u = 3$. We then discretize the continuous-time model using zero-order hold on the inputs with an appropriate sample time to obtain discrete-time linear models with state and input matrices of appropriate size (e.g., $A_{\scriptscriptstyle E} \in \mathbb{R}^{n_x \times n_x}, B_{\scriptscriptstyle E} \in \mathbb{R}^{n_x \times n_u}$).
\begin{alignat}{10}
        &x_{\scriptscriptstyle E}&&(h+1) &&= A_{\scriptscriptstyle E} &&x_{\scriptscriptstyle E}&&(h) &&+ B_{\scriptscriptstyle E} &&u_{\scriptscriptstyle E}(h) \\
  	&x_{\scriptscriptstyle I_i}&&(h+1) &&= A_{\scriptscriptstyle I} &&x_{\scriptscriptstyle I_i}&&(h) &&+ B_{\scriptscriptstyle I} &&u_{\scriptscriptstyle I_i}(h) \quad &&\forall \ &&i &&= 1, \ldots, M \\
	&x_{\scriptscriptstyle T_j}&&(h+1) &&= A_{\scriptscriptstyle T} &&x_{\scriptscriptstyle T_j}&&(h) &&+ B_{\scriptscriptstyle T} &&u_{\scriptscriptstyle T_j}(h) \quad &&\forall \ &&j &&= 1, \ldots, N
\end{alignat}

\subsubsection{Multi-Agent Model}

Consider an engagement between a subset of interceptors $\mathbb{I}_k \in \mathbf{I}$ and a threat $T_j \in \mathbf{T}$. Let us define $m:=\lvert \mathbb{I}_k \rvert$. Then the total number of agents, including the asset, is $q:=m+2$.
As a convention used in this paper, we build the augmented state vector for the $\{E, \mathbb{I}_k, T_j\}$ engagement by stacking the asset first, followed by the interceptors and then the threat:
\begin{align} \label{augmentedstates}
x := 
\left[ 
\arraycolsep=0.1pt\def\arraystretch{0.9}
\begin{array}{c}
 \begin{array}{c}
         x_{\scriptscriptstyle E} \\
  \end{array} \\
  \left[
  \arraycolsep=1.0pt\def\arraystretch{1.2}
  \begin{array}{c}
                        \phantom{x_{\scriptscriptstyle \mathbf{T}_l} } \\ 
                        x_{\scriptscriptstyle \mathbb{I}_k}  \\
                        \phantom{x_{\scriptscriptstyle \mathbf{T}_l} } \\
  \end{array}\right] \\
	x_{\scriptscriptstyle T_j} 
\end{array}
\right]  
= 
\left[ 
\arraycolsep=0.1pt\def\arraystretch{0.9}
\begin{array}{c}
 \begin{array}{c}
         x_{\scriptscriptstyle E} \\
  \end{array} \\
  \left[
  \arraycolsep=1.0pt\def\arraystretch{01.0}
  \begin{array}{c}
                        x_{\scriptscriptstyle \mathbb{I}_k [1]} \\ 
                        \vdots \\
                        x_{\scriptscriptstyle \mathbb{I}_k [m]} 
  \end{array}
  \right] \\
	         x_{\scriptscriptstyle T_j} \\
\end{array}
\right] \ .
\end{align}

We then stack the input vectors of the asset and the interceptors together into a single vector and the inputs of the threat in another. The asset and interceptors constitute a team driven by a collaborative guidance law while the threat is guided under a separate law that targets the asset while avoiding the interceptors.
\begin{align} \label{augmentedinputs}
u := 
\left[ 
\arraycolsep=0.1pt\def\arraystretch{0.9}
\begin{array}{c}
 \begin{array}{c}
         u_{\scriptscriptstyle E} \\
  \end{array} \\
  \left[
  \arraycolsep=0.1pt\def\arraystretch{1.2}
  \begin{array}{c}
                        \phantom{u_{\scriptscriptstyle I_1}} \\ 
                        u_{\scriptscriptstyle \mathbb{I}_k} \\
                        \phantom{u_{\scriptscriptstyle I_m}} \\
                        \end{array}\right]
  \end{array}
  \right]
=
\left[ 
\arraycolsep=0.1pt\def\arraystretch{0.9}
\begin{array}{c}
 \begin{array}{c}
         u_{\scriptscriptstyle E} \\
  \end{array} \\
  \left[
  \arraycolsep=0.1pt\def\arraystretch{0.9}
  \begin{array}{c}
                        u_{\scriptscriptstyle \mathbb{I}_k[1]} \\ 
                        \vdots \\
                        u_{\scriptscriptstyle \mathbb{I}_k[m]} \\
                        \end{array}\right]
  \end{array}\right] \ ,
\hspace{0.5cm} 
v := 
  \left[
  \arraycolsep=0.1pt\def\arraystretch{0.9}
  \begin{array}{c}
                        u_{\scriptscriptstyle T_j} 
  \end{array}\right] \ . 
\end{align}

The full multi-agent system can be expressed as
\begin{align} \label{gamemodel}
  	x(h+1) &= A x(h) + B_u u(h) + B_v v(h) \ ,
\end{align}
where the dimensions are listed below.
\begin{description}
  \item [$x \ \in \mathbb{R}^{q \cdot n_x \times 1}$]  \hspace{2.0cm} combined state  
  \item [$u \ \in \mathbb{R}^{(m +1) \cdot n_u  \times 1}$] \hspace{2.0cm} control of interceptors \& asset
  \item [$v \ \in \mathbb{R}^{n_u  \times 1}$] \hspace{2.0cm} control of threat 
  \item [$A \ \in \mathbb{R}^{q \cdot n_x \times q \cdot n_x}$] \hspace{2.0cm} state matrix of the game
  \item [$B_u \in \mathbb{R}^{q \cdot n_x \times (m+1) \cdot n_u}$] \hspace{2.0cm} input matrix of interceptors \& asset
  \item [$B_v \in \mathbb{R}^{q  \cdot n_x \times n_u}$] \hspace{2.0cm} input matrix of threat \\
\end{description}

The state and input matrices are constructed with the following MATLAB pseudocode:
\begin{align}
A     &= \mcode{blkdiag}(A_{\scriptscriptstyle E} \ , \ \mathbf{1}^{m} \otimes A_{\scriptscriptstyle I} \ , \ A_{\scriptscriptstyle T}) \label{A} \\
B_u &= \mcode{blkdiag}(B_{\scriptscriptstyle E} \ , \  [(\mathbf{1}^{m} \otimes B_{\scriptscriptstyle I}); \mathbf{0}^{n_x \times m \cdot n_u}]) \label{Bu} \\
B_v &=  [ \mathbf{0}^{(m+1) \cdot n_x \times n_u} \ ; \ B_{\scriptscriptstyle T} ] \label{Bv}
\end{align}

where \mcode{blkdiag} is the block diagonal concatenation of the input arguments and $\otimes$ is the Kronecker tensor product.

The matrices take the following form:
\begin{align}
A = \left[ 
\arraycolsep=1.0pt\def\arraystretch{1.0}
\begin{array}{c@{}c@{}c}
 \left[\begin{array}{c}
         A_{\scriptscriptstyle E} \\
  \end{array}\right] & \mathbf{0}^{n_x \times m \cdot n_x} & \mathbf{0}^{n_x \times n_x} \\
  \mathbf{0}^{m \cdot n_x \times n_x} & \left[\begin{array}{ccc}
                       A_{\scriptscriptstyle I} & \phantom{ A} &  \phantom{ A_{\scriptscriptstyle I_1}} \\ 
                       \phantom{ A_{\scriptscriptstyle I_1}} & \ddots &  \phantom{ A_{\scriptscriptstyle I_1}} \\
                       \phantom{ A_{\scriptscriptstyle I_1}} &  \phantom{ A} & A_{\scriptscriptstyle I} \\
                      \end{array}\right] & \mathbf{0}^{m \cdot n_x \times n_x} \\
\mathbf{0}^{n_x \times n_x} & \mathbf{0}^{n_x \times m \cdot n_x} & \left[ \begin{array}{ccc}
	         A_{\scriptscriptstyle T} 	
                      \end{array}\right]
\end{array}\right]
\end{align}

\begin{align}
B_u = \left[ 
\arraycolsep=0.1pt\def\arraystretch{1.0}
\begin{array}{c@{}c}
 \left[
 \arraycolsep=0.1pt\def\arraystretch{1.0}
 \begin{array}{c}
         B_{\scriptscriptstyle E} \\
  \end{array}\right] & \mathbf{0}^{n_x \times m \cdot n_u} \\
  \mathbf{0}^{m \cdot n_x \times n_u } & \left[ \begin{array}{ccc}
                       B_{\scriptscriptstyle I} & \phantom{ B} &  \phantom{ B_{\scriptscriptstyle I_1}} \\ 
                       \phantom{ B_{\scriptscriptstyle I_1}} & \ddots &  \phantom{ B_{\scriptscriptstyle I_1}} \\
                       \phantom{ B_{\scriptscriptstyle I_1}} &  \phantom{ B} & B_{\scriptscriptstyle I} \\
                      \end{array}\right] \\
   \mathbf{0}^{n_x \times n_u} & \mathbf{0}^{n_x \times m \cdot n_u} \\
\end{array}\right] \ , \hspace{0.3cm}
B_v = \left[ 
\arraycolsep=0.1pt\def\arraystretch{1.0}
\begin{array}{c}
 \begin{array}{c}
         \mathbf{0}^{n_x \times n_u} \\
  \end{array} \\
  \begin{array}{c}
                        \phantom{ B}  \\ 
                        \mathbf{0}^{m \cdot n_x \times n_u}  \\
                        \phantom{ B}
                        \end{array} \\
  \left[
   \arraycolsep=0.1pt\def\arraystretch{1.0}
  \begin{array}{ccc}
                        B_{\scriptscriptstyle T} 
                       \end{array}\right]
\end{array}\right] 
\end{align}
\\

\subsection{Collaborative guidance law}

In this section, we describe how we use results from finite-horizon, discrete-time, linear-quadratic dynamic games to induce collaborative guidance among the interceptors.
\vspace{0.1cm}

\subsubsection{Linear Quadratic Dynamic Game}

Following the work of \cite{Pachter}, we consider the linear, discrete-time dynamical system in equation (\ref{gamemodel}) where $u$ is the minimizing team and $v$ is the maximizing agent with respect to the following quadratic objective:
 \begin{align} \label{gamefunctional}
  	&J \left( \left\{ u(h) \right\}^{H-1}_{h=0} \ , \ \left\{ v(h) \right\}^{H-1}_{h=0} \ ; \ x(0)  \right) =  \frac{1}{2} x^\top(H) Q_{\scriptscriptstyle H} x(H) \ + \nonumber \\
	&\hspace{0.3cm} \frac{1}{2} \sum^{H-1}_{k=0} \left( x^\top(h) Q x(h) + u^\top (h) R_{\scriptstyle u} u(h) - v^\top(h) R_{\scriptstyle v} v(h) \right)
  \end{align}

with symmetric weight matrices and horizon denoted by

  \begin{description}
  \item [$Q_{\scriptscriptstyle H} \in \mathbb{R}^{q \cdot n_x \times q \cdot n_x}$] \hspace{2.7cm} terminal state weights
  \item [$Q\phantom{_H} \in \mathbb{R}^{q \cdot n_x \times q \cdot n_x}$] \hspace{2.7cm} running state weights
  \item [$R_{\scriptstyle u} \ \in \mathbb{R}^{(m +1) \cdot n_u \times (m+1) \cdot n_u}$]  \hspace{2.7cm} interceptor input weights
  \item [$R_{\scriptstyle v} \ \in \mathbb{R}^{n_u \times n_u}$] \hspace{2.7cm} threat input weights
  \item [$H \hspace{0.2cm} \in \mathbb{Z}_{\scriptscriptstyle \geq 0}$] \hspace{2.7cm} finite horizon.
\end{description}

In this two-player game, where one player consists of the interceptors and asset and the other player is the threat, each is aware of the multi-agent system model  (\ref{gamemodel}). Both players also have access to the system state (\ref{augmentedstates}) at any given time.
\vspace{1.0cm}

\begin{strip}
\medskip
\hrule
\begin{align}
F (h) &= \left( \mathbf{1}^{\scriptscriptstyle (m+1)\cdot n_u} - \left[ B_u^\top P(h+1) B_u + R_u \right]^{-1} \left[ B_u^\top P(h+1) B_v \right] \left[ B_v^\top P(h+1) B_v - R_v \right]^{-1} \left[ B_v^\top P(h+1) B_u \right] \right)^{-1}  \cdot \label{Fmat} \\
&\hspace{2.0cm} \left( \left[ B_u^\top P(h+1) B_u + R_u \right]^{-1} \left[B_u^\top P(h+1) B_v \right] \left[ B_v^\top P(h+1) B_v - R_v \right]^{-1} B_v^\top P(h+1) - \mathbf{1}^{\scriptscriptstyle q} \right) A \nonumber \\
G (h) &= \left( \mathbf{1}^{\scriptscriptstyle (n)\cdot n_u} -  \left[ B_v^\top P(h+1) B_v - R_v \right]^{-1} \left[ B_v^\top P(h+1) B_u \right] \left[ B_u^\top P(h+1) B_u + R_u \right]^{-1} \left[ B_u^\top P(h+1) B_v \right] \right)^{-1}  \cdot \label{Gmat} \\
&\hspace{1.75cm}  \left( \left[ B_v^\top P(h+1) B_v - R_v \right]^{-1} \left[B_v^\top P(h+1) B_u \right] \left[ B_u^\top P(h+1) B_u + R_u \right]^{-1} B_u^\top P(h+1) - \mathbf{1}^{\scriptscriptstyle q} \right) A \nonumber \\
P(h)  &= \Big( A + B_u F (h) + B_v G (h) \Big)^\top P(h+1) \Big( A + B_u F (h) + B_v G (h) \Big) + \left( Q + F^\top(h) R_u F(h) - G^\top (h)  R_v  G (h) \right) \label{Pmat} \\
%& \hspace{5.5cm} \text{for} \quad h = 0, \ldots, H-1 \nonumber \\
P(H) &= Q_{\scriptscriptstyle H} \label{Pfmat}
\end{align}
\hrule
\end{strip}

Under the following conditions:
\begin{align} 
\left[ B_u^T P(h+1) B_u + R_{\scriptstyle u} \right] &\succ 0 \hspace{0.6cm} \forall \ \ h = 0, \ldots, H-1 \label{condition1}\\
\left[ B_v^T P(h+1) B_v - R_{\scriptstyle v} \right] &\prec 0 \hspace{0.6cm} \forall \ \ h = 0, \ldots, H-1 \label{condition2}
\end{align}
it can be shown there exists a unique, closed-form solution to the discrete-time, linear-quadratic dynamic game (LQDG) described by equations (\ref{gamemodel}) and (\ref{gamefunctional}). More specifically, we have explicit formulas for the optimal guidance strategy of each player in the form of linear, state-feedback laws:
\begin{align}
u^* (h) = F(h) x(h) \hspace{0.6cm} \forall \ \ h = 0, \ldots, H-1 \\
v^* (h) = G(h) x(h) \hspace{0.6cm} \forall \ \ h = 0, \ldots, H-1
 \end{align}
where expressions for the time-varying, state-feedback gain matrices \iffalse $F:\mathbb{Z}_{\scriptscriptstyle \geq 0} \mapsto \mathbb{R}^{\scriptscriptstyle (\lvert \mathbb{I}_k \rvert +1) \cdot n_u \times N \cdot n_x}$ and $G: \mathbb{Z}_{\scriptscriptstyle \geq 0} \mapsto \mathbb{R}^{\scriptscriptstyle \lvert \mathbb{T}_l \rvert \cdot n_u \times N \cdot n_x}$ \fi are shown in equation (\ref{Fmat}) for the minimizing player and in equation (\ref{Gmat}) for the maximizing player. \iffalse (We use the notation $\mathbf{1}^{\scriptscriptstyle N}$ to represent the $N \times N$ identity matrix.) \fi

Assuming that conditions (\ref{condition1}) and (\ref{condition2}) are satisfied, the gain matrices can be found by recursively solving the Riccati difference equation (\ref{Pmat}) from $h = H-1$ to  $h=0$ with equation (\ref{Pfmat}). Note that a family of state-feedback gains, computed with different weight matrices and horizon lengths, can be solved ahead of time and stored for later use in real-time collaborative guidance.
\vspace{0.1cm}

\subsubsection{LQDG weights}
Advanced methods for selecting or tuning LQR weight matrices exist, with a whole section devoted to the topic in \cite{Anderson}. Here we describe a simple approach to designing the LQDG weight matrices $Q$, $Q_{\scriptscriptstyle H}$, $R_{\scriptstyle u}$, $R_{\scriptstyle v}$ for an asset-guarding game. Since we are concerned with the terminal state of the finite-horizon game and not the trajectories taken to reach the terminal states, we take an approach that is similar to that used to derive the classical proportional navigation (PN) guidance law in \cite{Ho}. We penalize only the control effort and the relative terminal states. Hence, the running state weight matrix is 
\begin{align} \label{Q}
	Q = \mathbf{0}^{q \cdot n_x \times q \cdot n_x}
\end{align}

For the terminal state and input weight matrices, we use a rule of thumb \cite{Bryson}, often called Bryson's Rules \cite{Hespanhalinear}, to make the terms in the objective dimensionless and of the same order of magnitude. In addition to this normalization, we also introduce relative weighting for each state and control variable. To design the input weight matrices, we define block diagonal matrices $R_{\scriptstyle u}$ and $R_{\scriptstyle v}$ as 
\begin{align}
	R_{\scriptstyle u} &= \rho_u \cdot \mcode{blkdiag}(R_{\scriptscriptstyle E} \ , \ \mcode{kron}(\mcode{eye}(m), R_{\scriptscriptstyle I}) ) \label{Ru} \\
	R_{\scriptstyle v} &= \rho_v \cdot R_{\scriptscriptstyle T} \label{Rv}
\end{align}

The submatrices along the diagonal (i.e., $R_{\scriptscriptstyle E}$, $R_{\scriptscriptstyle I}$, $R_{\scriptscriptstyle T}$) normalize the inputs with respect to the horizon length $H$ and maximum acceptable input value $u_{\scriptscriptstyle max}$ :
\begin{align}
R_{\scriptscriptstyle E} = \ R_{\scriptscriptstyle I} &= \tfrac{\alpha}{H (u_{\scriptscriptstyle max})^2} \cdot \mathbf{1}^{\scriptscriptstyle n_u} \ \  \forall \hspace{0.2cm} i = 1, \ldots, m \label{Ri} \\
R_{\scriptscriptstyle T} &= \tfrac{\beta}{H (u_{\scriptscriptstyle max})^2} \cdot \mathbf{1}^{\scriptscriptstyle n_u}  \ , \label{Rt}
\end{align}
where we set $R_{\scriptscriptstyle E} = \ R_{\scriptscriptstyle I}$ for the example in this paper.
We use $\alpha := \tfrac{1}{(m +1)n_u}$ to scale each diagonal element of the $R_{\scriptstyle u} $ weight matrix so that their sum is equal to 1. Similarly, $\beta := \tfrac{1}{n \cdot n_u}$ is used to scale the diagonal elements of the $R_{\scriptstyle v}$ weight matrix. Finally, $\rho_u \text{ and } \rho_v \in \mathbb{R}$ weigh the input cost against the terminal state cost.

To design the terminal state weight matrix $Q_{\scriptscriptstyle H}$, we first introduce a relative state matrix $Q_{\scriptscriptstyle rel}$ illustrated below
\begin{align} \label{Qrel}
Q_{\scriptscriptstyle rel} := \left[ 
\begin{array}{c@{}c@{}c}
\begin{array}{c}
                      $-$\mathbf{S} \\
\end{array}
& \begin{array}{ccc}
                      \mathbf{O} & \ldots  & \mathbf{O}  \\ 
\end{array} 
& \begin{array}{c}
                      $\phantom{-}$\mathbf{S} \\ 
\end{array} \\
\begin{array}{c}
                      $\phantom{-}$ \mathbf{O} \\
                      $\phantom{-}$\vdots \\
                      $\phantom{-}$ \mathbf{O}
\end{array}
& \begin{array}{ccc}
                      \mathbf{S} & \phantom{\mathbf{0}}  & \phantom{\mathbf{0}} \\ 
                      \phantom{\mathbf{0}} & \ddots & \phantom{\mathbf{0}} \\
                      \phantom{\mathbf{0}} & \phantom{\mathbf{0}} & \mathbf{S}
\end{array}
& \begin{array}{c}
                      $-$\mathbf{S} \\ 
                      \vdots \\
                      $-$\mathbf{S}
\end{array}
\end{array}\right]
\end{align}
where $\mathbf{S}:= \left[ \mathbf{1}^{\scriptscriptstyle \sfrac{n_x}{2}},\mathbf{0}^{\scriptscriptstyle (\sfrac{n_x}{2}) \times (\sfrac{n_x}{2})} \right]$ and $\mathbf{O}:= \left[  \mathbf{0}^{\scriptscriptstyle (\sfrac{n_x}{2}) \times (n_x)} \right]$. The relative terminal positions can be expressed with the following matrix-vector multiplication:
\begin{align} \label{Qrelx}
Q_{\scriptscriptstyle rel} x(H) &=
\left[ 
\begin{array}{c@{}c@{}c}
\begin{array}{c}
                      $-$\mathbf{S} \\
\end{array}
& \begin{array}{ccc}
                      \mathbf{O} & \ldots  & \mathbf{O}  \\ 
\end{array} 
& \begin{array}{c}
                      $\phantom{-}$\mathbf{S} \\ 
\end{array} \\
\begin{array}{c}
                      $\phantom{-}$ \mathbf{O} \\
                      $\phantom{-}$\vdots \\
                      $\phantom{-}$ \mathbf{O}
\end{array}
& \begin{array}{ccc}
                      \mathbf{S} & \phantom{\mathbf{0}}  & \phantom{\mathbf{0}} \\ 
                      \phantom{\mathbf{0}} & \ddots & \phantom{\mathbf{0}} \\
                      \phantom{\mathbf{0}} & \phantom{\mathbf{0}} & \mathbf{S}
\end{array}
& \begin{array}{c}
                      $-$\mathbf{S} \\ 
                      \vdots \\
                      $-$\mathbf{S}
\end{array}
\end{array}\right]
\left[ 
\arraycolsep=0.1pt\def\arraystretch{0.9}
\begin{array}{c}
 \begin{array}{c}
         x_{\scriptscriptstyle E} \\
  \end{array} \\
  \left[
  \arraycolsep=1.0pt\def\arraystretch{01.0}
  \begin{array}{c}
                        x_{\scriptscriptstyle \mathbb{I}_k [1]}  \\ 
                        \vdots \\
                        x_{\scriptscriptstyle \mathbb{I}_k [m]} 
  \end{array}
  \right] \\
	         x_{\scriptscriptstyle T_j} \\
\end{array}
\right] (H) \nonumber \\
&= \left[
\begin{array}{c}
	              \ \ p_{\scriptscriptstyle T_j} (H) - p_{\scriptscriptstyle E} (H) \\
                      p_{\scriptscriptstyle \mathbb{I}_k [1]} (H) - p_{\scriptscriptstyle T_j} (H) \\
                      \ \ \vdots \\
                      p_{\scriptscriptstyle \mathbb{I}_k [m]} (H) - p_{\scriptscriptstyle T_j} (H) 
\end{array}
\right] \in \mathbb{R}^{(m+1) \cdot (\sfrac{n_x}{2})} \ ,
\end{align}
where the variable $p \in \mathbb{R}^{\scriptscriptstyle \sfrac{n_x}{2}}$ is used for the position vector. The first set of elements in (\ref{Qrelx}) is the threat's terminal position relative to the asset's terminal position. Each subsequent set contains, for the group $\mathbb{I}_k$, each interceptor's terminal position relative to that of the threat $T_j$. 

We introduce another intermediate matrix that is parameterized by a vector of positive parameters, $d \in \mathbb{R}^{m+1}$  :
\begin{align} \label{Qw}
	Q_{\scriptscriptstyle w}(d) := \mcode{blkdiag}\left(-w(d_{\scriptscriptstyle 1}), \ldots, w(d_{\scriptscriptstyle m+1})\right) 
\end{align}
where
\begin{align}
	w(z) := \mcode{diag}\left(\frac{z}{r_{max}^2},\frac{z}{r_{max}^2},\frac{z}{r_{max}^2}\right)
\end{align}
The constant $r_{\scriptscriptstyle max}$ is the maximum acceptable relative distance. Note that the first diagonal block of (\ref{Qw}) is negative and recall that the interceptors and asset team is the minimizer and the threat is the maximizer of the LQDG objective (\ref{gamefunctional}). Since the interceptors are attempting to maximize (threat is attempting to minimize) the final distance between the threat and the asset, we place a minus sign in front of this first diagonal block. For the remaining diagonal blocks we keep the sign positive because the interceptors are indeed attempting to minimize their relative distance to the threat, and the threat is trying to maximize its distance from all interceptors.
\iffalse
\begin{align}
	w(z) := \begin{bmatrix}
	\frac{z}{r_{max}^2} & \phantom{\tfrac{1}{2}} & \phantom{\tfrac{1}{2}} & \phantom{\tfrac{1}{2}} & \phantom{\tfrac{1}{2}} & \phantom{\tfrac{1}{2}} \\
	\phantom{\tfrac{1}{2}} & \frac{z}{r_{max}^2} & \phantom{\tfrac{1}{2}} & \phantom{\tfrac{1}{2}} & \phantom{\tfrac{1}{2}} & \phantom{\tfrac{1}{2}} \\
	\phantom{\tfrac{1}{2}} & \phantom{\tfrac{1}{2}} & \frac{z}{r_{max}^2} & \phantom{\tfrac{1}{2}} & \phantom{\tfrac{1}{2}} & \phantom{\tfrac{1}{2}} \\
	\phantom{\tfrac{1}{2}} & \phantom{\tfrac{1}{2}} & \phantom{\tfrac{1}{2}} & \frac{1}{V_{max}^2} & \phantom{\tfrac{1}{2}} & \phantom{\tfrac{1}{2}} \\
	\phantom{\tfrac{1}{2}} & \phantom{\tfrac{1}{2}} & \phantom{\tfrac{1}{2}} & \phantom{\tfrac{1}{2}} & \frac{1}{V_{max}^2} & \phantom{\tfrac{1}{2}} \\
	\phantom{\tfrac{1}{2}} & \phantom{\tfrac{1}{2}} & \phantom{\tfrac{1}{2}} & \phantom{\tfrac{1}{2}} & \phantom{\tfrac{1}{2}} & \frac{1}{V_{max}^2}
	\end{bmatrix}
\end{align}
\fi

We now define the terminal state weight matrix $Q_{\scriptscriptstyle H}$ as
\begin{align}
	Q_{\scriptscriptstyle H}(d) := Q_{\scriptscriptstyle rel}^\top Q_{\scriptscriptstyle w}(d) Q_{\scriptscriptstyle rel}  
\end{align}
so that the terminal cost may be expressed as:
\begin{align}
	&x^\top(H)  Q_{\scriptscriptstyle H}(d) x(H) = \nonumber \\
	&\tfrac{1}{r_{max}^2} \begin{bmatrix} \text{-}d_1 \\ \phantom{\text{-}}d_2 \\ \vdots \\ d_{m+1} \end{bmatrix}^\top \begin{bmatrix} \lVert p_{\scriptscriptstyle T_j} (H) \ - \ p_{\scriptscriptstyle E} (H) \rVert_{\scriptscriptstyle 2}^{\scriptscriptstyle 2} \\
	\lVert p_{\scriptscriptstyle \mathbb{I}_k [1]} (H) - p_{\scriptscriptstyle T_j} (H) \rVert_{\scriptscriptstyle 2}^{\scriptscriptstyle 2} \\ \vdots \\
	\lVert p_{\scriptscriptstyle \mathbb{I}_k [m]} (H) - p_{\scriptscriptstyle T_j} (H) \rVert_{\scriptscriptstyle 2}^{\scriptscriptstyle 2} \end{bmatrix}                      
\end{align}

We may constrain the parameter vector to belong to some finite set, $d \in \mathcal{D}$, or we may place lower and upper bounds on the values it may take: $\underline{d} \leq d \leq \overline{d}$. One simple approach of creating a finite space $\mathcal{D}$ to search over is to compose it with a set of parameter vectors that each make one element value much larger than the others. The corresponding relative distance will be penalized heavily. For example, if we make $d_2$ very large compared to the other parameters, we are essentially choosing interceptor $\mathbb{I}_k [1]$ to intercept the threat while remaining interceptors in the team, with less weight, support the effort by influencing the threat to evade them.

Now that we have described how we design the weight matrices $Q$, $Q_{\scriptscriptstyle H}$, $R_{\scriptstyle u}$, $R_{\scriptstyle v}$, we use the following expression
\begin{align} \label{LQDGcode}
\left[   \{F(h)\}_{\scriptscriptstyle h=0}^{\scriptscriptstyle H-1}, \{G(h)\}_{\scriptscriptstyle h=0}^{\scriptscriptstyle H-1} \right] \leftarrow \tt{LQDG \left( A, B_u, B_v, Q_{\scriptscriptstyle H} (d), Q, R_{\scriptstyle u} , R_{\scriptstyle v} \right)}
 \end{align}
to refer to the process of obtaining the solution to the LQDG, where $F$ and $G$ are the sequence of state-feedback gain matrices for the minimizer and maximizer, respectively.

We now take a step back and note that the LQDG is a zero-sum game with a single objective that is minimized by one team and maximized by the other. However, in an asset-guarding game we may assume that each team has a different objective. For example, the threat group is more concerned with capturing the asset than evading the interceptors. On the other hand, the interceptors may place more emphasis on intercepting the threats than thwarting the threats' effort to hit the asset. For this reason, we construct different objectives for the interceptors and threats, deriving guidance laws for each team based on its particular objective. We assume that the threat's terminal weight matrix $Q_G$ is known, that it will place heavy weight on its relative distance to the asset compared to those with the interceptors.
\begin{align}
&\left[   \{F(h)\}_{\scriptscriptstyle h=0}^{\scriptscriptstyle H-1}, \sim \right] \leftarrow \tt{LQDG \left( A, B_u, B_v, Q_{\scriptscriptstyle F} (d), Q, R_{\scriptstyle u} , R_{\scriptstyle v} \right)} \\
&\left[  \sim, \{G(h)\}_{\scriptscriptstyle h=0}^{\scriptscriptstyle H-1} \right] \leftarrow \tt{LQDG \left( A, B_u, B_v, Q_{\scriptscriptstyle G} \phantom{(d)}, Q, R_{\scriptstyle u} , R_{\scriptstyle v} \right)} 
\end{align}
We note that the guidance law for a team is only optimal if the opposing team is also acting optimally with respect to the same objective. Hence, the guidance laws resulting from this approach are sub-optimal. Nevertheless, interesting cooperative team behavior may be observed.

\subsection{Reward definition}

Recall that the reward matrix $r$ in the objective of the assignment problem (\ref{cost}) must be computed prior to solving the problem. For every combination of interceptor group $\mathbb{I}_k$ for $k=1,\ldots,2^M-1$ and threats $T_j$, the corresponding element of the reward matrix is computed with
\begin{align}
	r_{jk} := \mcode{solveReward}(T_j, \mathbb{I}_k) \ ,
\end{align}

where we define the function \mcode{solveReward} as
\begin{align}
  &\mcode{solveReward}(T_j, \mathbb{I}_k) := \nonumber \\
  &\max_{d \in \mathcal{D}, \ H \in \mathcal{H}} \quad \lVert p_{\scriptscriptstyle T_j} (H) - p_{\scriptscriptstyle E} (H) \rVert_{\scriptscriptstyle 2}  \label{rkl} \\
  &\text{subject to} \nonumber \\
  &\exists \ I_i \in \mathbb{I}_k \ \ni \ \lVert p_{\scriptscriptstyle I_i} (H) - p_{\scriptscriptstyle T_j} (H) \rVert_{\scriptscriptstyle 2} \leq r_c \label{constraint6a}  \\
  &x(0) = [x_{\scriptscriptstyle E}^\top (0), x_{\scriptscriptstyle \mathbb{I}_k}^\top (0), x_{\scriptscriptstyle T_j}^\top (0) ]^\top \label{constraint0a} \\
  &x (h+1) = A x(h) + B_u u (h) + B_v v (h)   \label{constraint1a} \\
  &u (h) = sat \left( F(h) x(h), u_{max} \right) \label{constraint2a} \\
  &v (h) = sat \left( G(h) x(h), v_{max} \right) \label{constraint3a}\\
  & \{F(h)\}_{\scriptscriptstyle h=0}^{\scriptscriptstyle H-1} \leftarrow \tt{LQDG \left( A, B_u, B_v, Q_{\scriptscriptstyle F}(d) , Q, R_{\scriptstyle u} , R_{\scriptstyle v}, H \right)} \label{constraint4a} \\
  & \{G(h)\}_{\scriptscriptstyle h=0}^{\scriptscriptstyle H-1} \leftarrow \tt{LQDG \left( A, B_u, B_v, Q_{\scriptscriptstyle G}\phantom{d} , Q, R_{\scriptstyle u} , R_{\scriptstyle v}, H \right)} \label{constraint5a}
\end{align}
where $p_{\scriptscriptstyle I_i} (T)$ and $p_{\scriptscriptstyle T_j} (T) \in \mathbb{R}^3$ refer to the \textit{predicted} positions of interceptor $I_i$ and threat $T_j$ in inertial space at some terminal time $T$ and $r_c \in \mathbb{R}$ represents a desired capture (or lethal) radius. The terminal positions of each interceptor and threat are predicted using equation (\ref{constraint1a}) where the state feedback commands are computed with  (\ref{constraint4a}) and (\ref{constraint5a}), and  elementwise saturated with respect to some maximum acceleration value assumed for each agent in equations (\ref{constraint2a}) and (\ref{constraint3a}). If there does not exist a terminal time $H \in \mathcal{H}$ such that the intercept condition in (\ref{constraint6a}) is satisfied, then we set the value of $r_{jk}$ to a large negative value.

\section{SIMULATION AND RESULTS}

\begin{figure}[h]
\centering
\includegraphics[width=.4\textwidth]{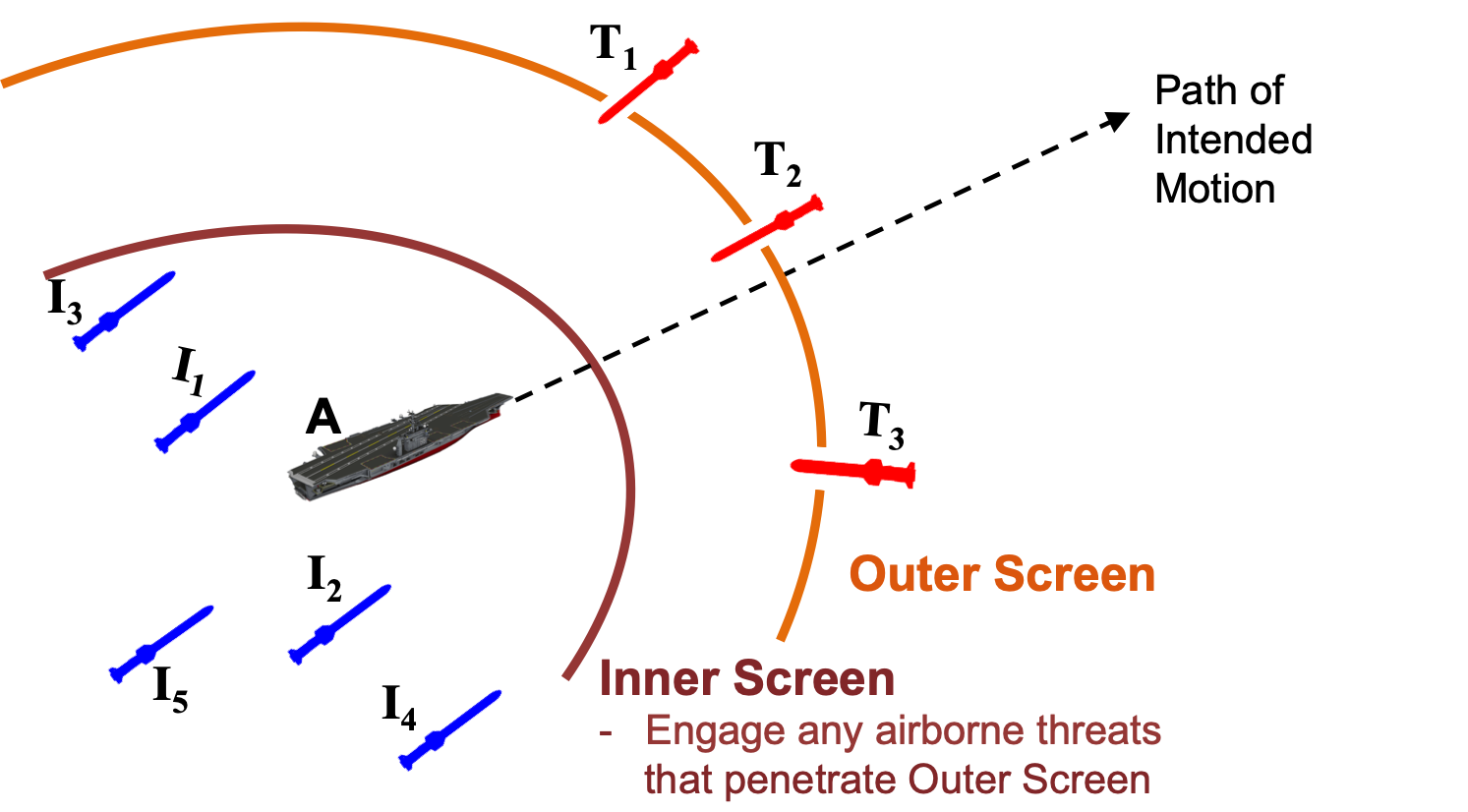}
\caption{Asset-guarding engagement with five interceptor missiles against three threat missiles}
\label{fig:example}
\end{figure}

We consider the scenario depicted in figure (\ref{fig:example}) where three threat missiles have penetrated an outer defense screen to reach within a 8000ft radius of a high-value asset, each with a speed between 2400 and 2800ft/s. Five interceptor missiles within a 500ft radius of the asset are launched with initial speeds between 1800 and 2200ft/s. The maximum accelerations of the threats and interceptors are assumed to be 30Gs and 20Gs, respectively. Although not hypersonic, the threat missiles are traveling faster and are more maneuverable than the interceptors. The asset is traveling at a constant velocity of 50ft/s and is assumed to be non-maneuverable compared to speeds and accelerations of the missiles.

The objective of the interceptors is to intercept ($r_c \leq 20$ft) all threat missiles at distances as far as possible from the asset. The interceptors are optimally assigned to the threats and controlled by a collaborative intercept guidance law. In this example, we assume that the threat missiles do not collaborate with each other. However, each threat missile is capable of tracking the positions and velocities of collaborative interceptor groups that are targeting it. Hence, the threats may perform evasive maneuvers to avoid the interceptors while on trajectories to hit the asset.

For our 6 degree-of-freedom (DOF) simulation, we use a non-linear, coupled 3-axis generic missile airframe model from \cite{Tan}. Six state variables (missile speed, angle-of-attack, sideslip angle, roll rate, pitch rate, and yaw rate) are used to describe missile motion in the body frame. Aerodynamic forces and moments that enter the equations of motion are modeled in terms of benchmark aero-coefficients that are also provided in \cite{Tan}. Another six states are used to describe the position and attitude of the missile with respect to an inertial reference frame. The missile is actuated through controls surfaces in the roll, pitch, and yaw channels. We assume that the missile has access to accurate state information. The simulation model includes subsystems for the guidance law, autopilot, and fin actuators as described in \cite{Zarchan}. \iffalse In addition to the collaborative LQDG guidance law, a 3-dimensional, true proportional navigation law based on \cite{Palumbo1} and \cite{Palumbo2} is included in the simulation environment.\fi A traditional 3-loop autopilot architecture, as detailed in \cite{Mracek1}, is implemented for both the pitch and yaw channels whereas a simple proportional-integral controller is used for the roll channel. The autopilot subsystem is gain-scheduled to obtain suitable performance throughout the operational flight envelope of the missiles. Finally, a first-order model is used for the fin actuators.

\subsection{Solution to Assignment Problem}

To compute the reward matrix for our assignment problem, we first design the LQDG feedback gains in (\ref{constraint4a}) and (\ref{constraint5a}) for a set of horizon values $\mathcal{H}$, a coarse parameter search space $\mathcal{D}$,  $\rho_u = \rho_v= 1000$, $u_{\scriptscriptstyle max} = 20$G, $v_{\scriptscriptstyle max} = 30$G, and $r_{\scriptscriptstyle max}=20$ft. For each interceptor, threat and asset we use initial conditions that satisfy the scenario description.

Using a computer with a 2.7 GHz Intel Core i5 64-bit processor and 8 GB of RAM, a computation time of 52.6 seconds was required to construct the reward matrix and 0.9 seconds was required to solve the assignment problem. With parallel computing, we expect that the reward matrix computation time can be reduced significantly. 

In table (\ref{tab:2}) we list the optimal assignments and note that the predicted time to intercept all threats is 1.91 seconds. We verify that each of the threats is predicted to enter the lethal radius of one of the interceptors ($\lVert \mathbf{p_{\scriptscriptstyle I_i}} (H) - \mathbf{p_{\scriptscriptstyle T_j }} (H)\rVert_{\scriptscriptstyle 2} < 20$ ft). We also observe that all threats are predicted to be intercepted at a distance of at least 2115.2ft from the asset ($\lVert \mathbf{p_{\scriptscriptstyle T_j}} (H) - \mathbf{p_{\scriptscriptstyle A }} (H)\rVert_{\scriptscriptstyle 2}$).

\begin{table}
\begin{center}
\begin{tabular}{SSSS} 
 \toprule
     {Assignment} & {$H$} &  {\begin{tabular}{@{}c@{}}{ \scriptsize $\lVert \mathbf{p_{\scriptscriptstyle I_i}} (H) - \mathbf{p_{\scriptscriptstyle T_j }} (H)\rVert_{\scriptscriptstyle 2} $} \\ {\scriptsize \color{gray}{Actual (Predicted)}}\end{tabular}}   &  {\begin{tabular}{@{}c@{}}{\scriptsize $\lVert \mathbf{p_{\scriptscriptstyle T_j}} (H) - \mathbf{p_{\scriptscriptstyle A }} (H)\rVert_{\scriptscriptstyle 2} $} \\ {\scriptsize \color{gray}{Actual (Predicted)}}\end{tabular}}  \\ \midrule
     {$\{ \mathbf{I_4} \} \rightarrow \{ \mathbf{T_1} \}$} & {1.65s} & {19.2 (17.8) ft} & {2145.2 (2145.5) ft} \\ \midrule
     {$\{ \mathbf{I_1} \} \rightarrow \{ \mathbf{T_2} \}$} & {1.63s} & {19.4 (14.5) ft} & {2138.9 (2115.2) ft}  \\ \midrule
     {$\{ I_2,I_3,\mathbf{I_5} \} \rightarrow \{ \mathbf{T_3} \}$} & {1.91s} & {14.7 (19.3) ft} & {2602.4 (2621.3) ft}  \\ \bottomrule
\end{tabular} 
\end{center}
\caption{Assignments and relative terminal distances for both predicted and 6DOF simulation trajectories} \label{tab:2}
\end{table}

\subsection{6DOF Simulation Results}

\begin{figure}[h]
\centering
\includegraphics[width=.49\textwidth]{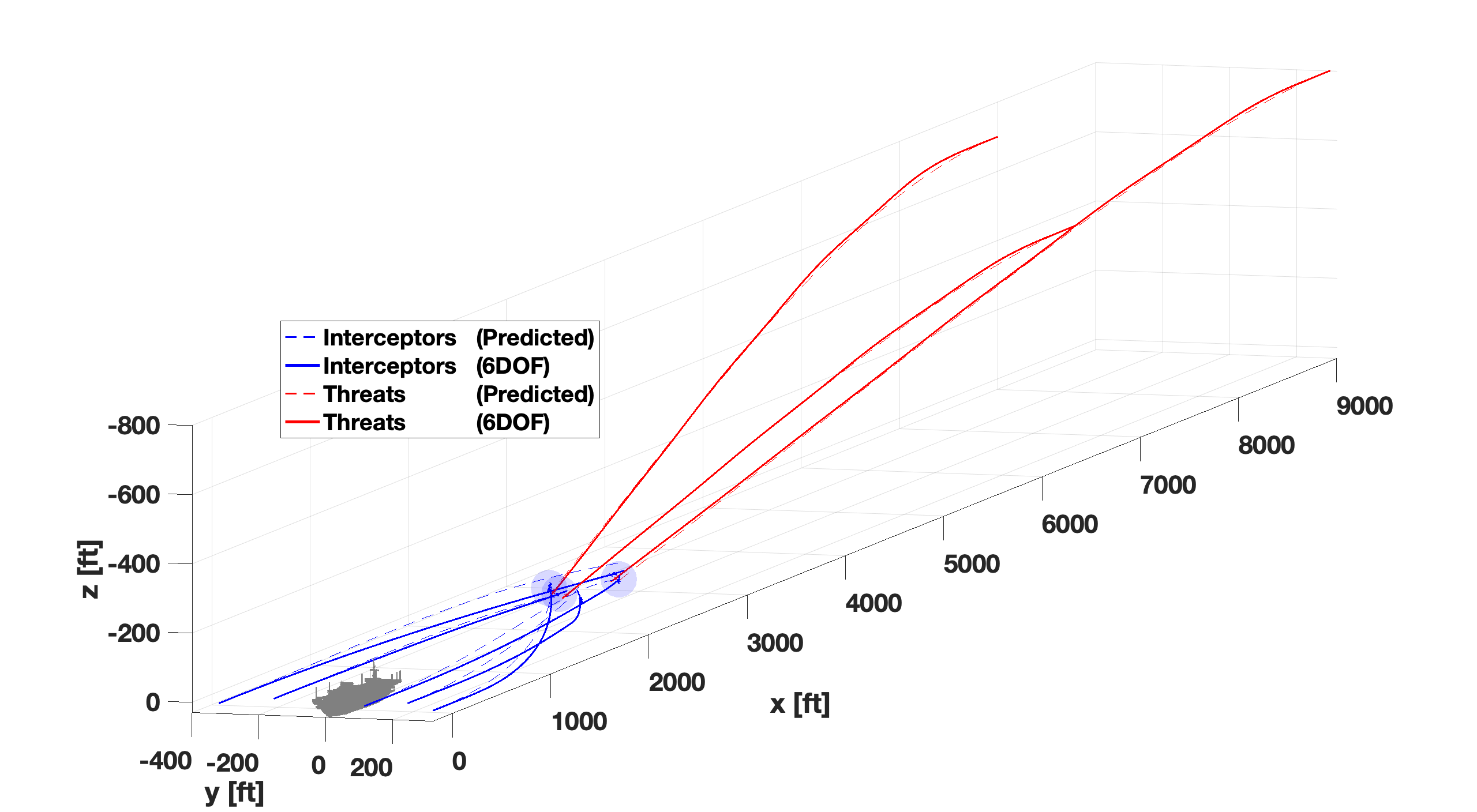}
\caption{Predicted \& 6DOF missile trajectories of (blue) interceptors, (red) threats, and aircraft carrier as (gray) asset}
\label{fig: zoomout}
\end{figure}

\begin{figure}[h]
\centering
\includegraphics[width=.49\textwidth]{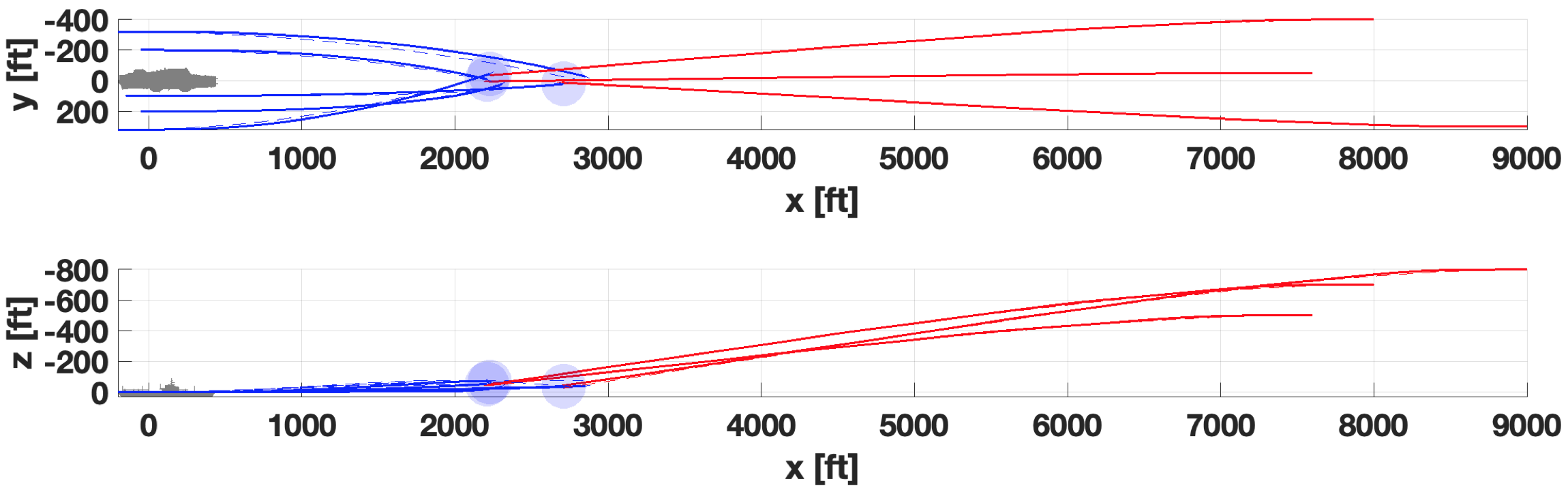}
\caption{Top (birds-eye) and side views of engagement}
\label{fig: topside}
\end{figure}

When we apply the assignment and corresponding LQDG guidance laws to the 6DOF simulation we observe that all threats are intercepted within the predicted horizons, as shown in table (\ref{tab:2}). With properly-tuned autopilots, we note that the 6DOF simulation results in relative terminal distances that match well with the predictions. In figures (\ref{fig: zoomout}) and (\ref{fig: topside}), we overlay the 6DOF missile trajectories on top of the predicted trajectories. We observe that the autopilot-controlled 6DOF missile system can effectively track the guidance commands used in the predicted trajectories. The blue bubbles represent the lethal radius of interceptors, signifying successful intercept of threat missiles.

\begin{figure}[h]
\centering
\includegraphics[width=.49\textwidth]{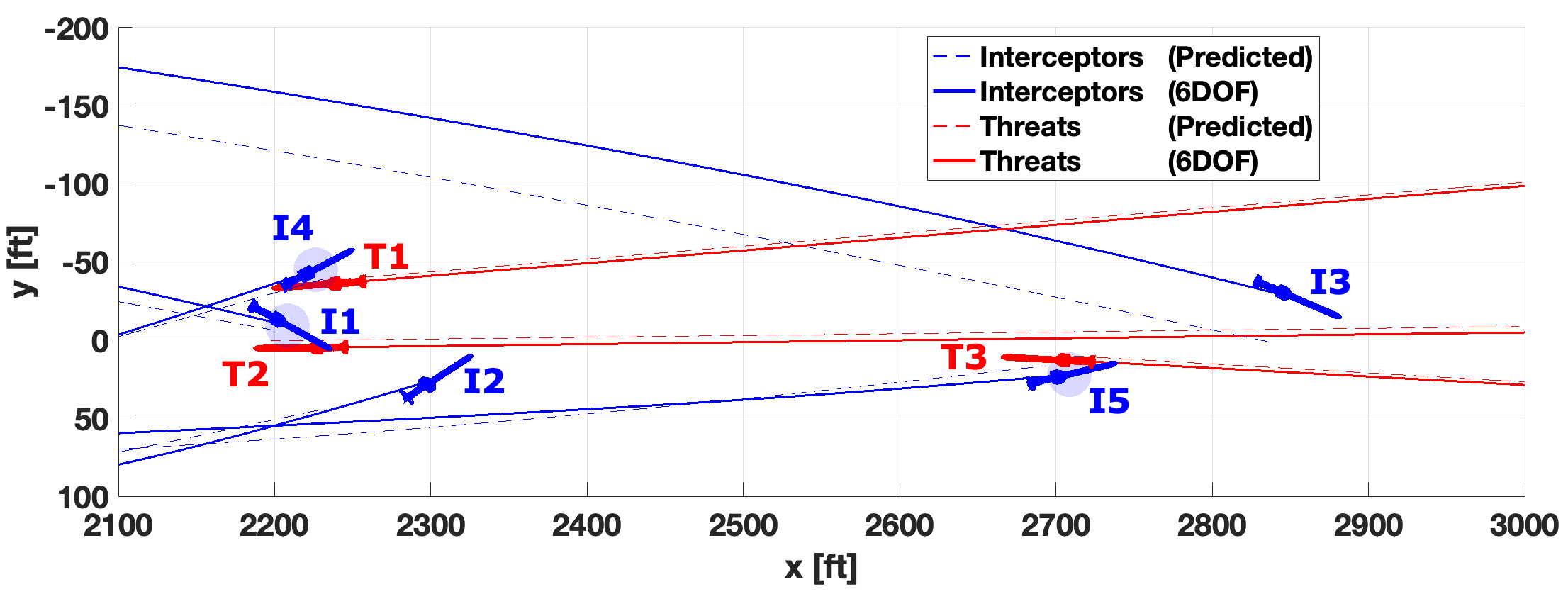}
\caption{Zoomed-in, birds-eye view of engagement endgame}
\label{fig: zoomin}
\end{figure}

In figure (\ref{fig: zoomin}) we show a zoomed-in, birds-eye view of the engagement endgame (i.e., the last few miliseconds of the engagement). We observe the one-on-one intercepts of $\{ I_4 \} \rightarrow \{ T_1 \}$ and $\{ I_1 \} \rightarrow \{ T_2 \}$. The sub-engagement of interest is that of $\{ I_2,I_3,1_5 \} \rightarrow \{ T_3 \}$. Threat $T_3$, which has the greatest speed of the threats, is coralled into the trajectory of $I_5$ by $I_2$ and $I_3$. Here we can clearly see the effect of the collaborative guidance law where the interceptor with a strategic position and velocity advantage is employed to intercept while the other interceptors influence the threat's trajectory by forcing it to evade them.

\section{CONCLUSIONS}

We approach the multi-body asset-guarding game by decomposing it into two steps: predicting multi-body trajectories using LQDG as a collaborative guidance law, and using the predictions to solve an assignment problem. Through simulation, we observe that our approach can provide assignments and guidance laws that perform well despite the linear assumptions made in the approach.


%%%%%%%%%%%%%%%%%%%%%%%%%%%%%%%%%%%%%%%%%%%%%%%%%%%%%%%%%%%%%%%%%%%%%%%%%%%%%%%%

%%%%%%%%%%%%%%%%%%%%%%%%%%%%%%%%%%%%%%%%%%%%%%%%%%%%%%%%%%%%%%%%%%%%%%%%%%%%%%%%

\iffalse
%%%%%%%%%%%%%%%%%%%%%%%%%%%%%%%%%%%%%%%%%%%%%%%%%%%%%%%%%%%%%%%%%%%%%%%%%%%%%%%%
\section*{APPENDIX}

Appendixes should appear before the acknowledgment.
\fi

\section*{ACKNOWLEDGMENT}

We thank Jyot Buch for work on our MATLAB Simlunk\textsuperscript{\textregistered} environment, Kate Schweidel for work on the autopilot implementation and He Yin for helpful discussions in writing the paper. The authors gratefully acknowledge support from the Office of Naval Research under grant N00014-18-1-2209. Andrew Packard acknowledges the generous support from the FANUC Corporation.\\

%%%%%%%%%%%%%%%%%%%%%%%%%%%%%%%%%%%%%%%%%%%%%%%%%%%%%%%%%%%%%%%%%%%%%%%%%%%%%%%%


\begin{thebibliography}{99}

\bibitem{MDA} U.S. Department of Defense, Missile Defense Agency, ''2019 Missile Defense Review,'' \url{https://www.mda.mil/news/downloadable_resources.html}.
%%% Differential Games
\bibitem{Isaacs} R. Isaacs, ``Differential Games'', New York, NY, USA: John Wiley and Sons, 1965.

%%% Dynamic game theory
\bibitem{Basar1} T. Ba\c sar, G.J. Olsder, ``Dynamic Noncooperative Game Theory'', New York, NY, USA: Academic Press, 1995.
\bibitem{Engwerda} J.C. Engwerda, ``LQ Dynamic Optimization and Differential Games'', West Sussex, England: John Wiley and Sons, 2005.
\bibitem{Hespanhagame} J.P. Hespanha, ``Noncooperative Game Theory'', Princeton, NJ, USA: Princeton University Press, 2017.
%\bibitem{Basar2} T. Ba\c sar, P. Bernhard, ``H$^{\infty}$- Optimal Control and Related Minimax Design Problems'', Boston, MA, USA: Birkh\" auser, 1995.

%%% Collaborative guidance laws
%%% Pachter papers
\bibitem{Garcia1} E. Garcia, D. W. Casbeer, K. Pham, M. Pachter, ``Cooperative Aircraft Defense from an Attacking Missile,'' in \textit{Proceedings of the 2014 IEEE Conference on Decision and Control}, Los Angeles, California, USA, December 2014.
\bibitem{Garcia2} E. Garcia, D. W. Casbeer, K. Pham, M. Pachter,  ``Cooperative Aircraft Defense from an Attacking Missile using Proportional Navigation,'' in \textit{Proceedings of the 2015 AIAA Guidance, Navigation, and Control Conference}, Kissimmee, Florida, USA, January 2015.
\bibitem{Garcia3} E. Garcia, D. W. Casbeer, M. Pachter, ``Cooperative Strategies for Optimal Aircraft Defense from an Attacking Missile,'' \textit{Journal of Guidance, Control, and Dynamics}, vol. 38, no. 8, pp. 1510-1520.
%\bibitem{Ratnoo}  A. Ratnoo, T. Shima,  ``Guidance Strategies Against Defended Aerial Targets,'' \textit{Journal of Guidance, Control, and Dynamics}, vol. 35, no. 4, pp. 1059-1068, August 2012.


\bibitem{Prokopov} O. Prokopov, T. Shima, ``Linear Quadratic Optimal Cooperative Strategies for Active Aircraft Protection,'' \textit{Journal of Guidance, Control, and Dynamics}, vol. 36, no. 3, pp. 753-764, June 2013.

%%% Shima papers
\bibitem{Shaferman1} V. Shaferman, T. Shima, ``Cooperative Multiple Model Adaptive Guidance for an Aircraft Defending Missile,'' \textit{Journal of Guidance, Control, and Dynamics}, vol. 33, no. 6, pp. 1801-1813, December 2010.

\bibitem{Shima} T. Shima, ``Optimal Cooperative Pursuit and Evasion Strategies Against a Homing Missile,'' \textit{Journal of Guidance, Control, and Dynamics}, vol. 34, no. 2, pp. 414-425, April 2011.

\bibitem{Perelman} A. Perelman, T. Shima, I. Rusnak, ``Cooperative Differential Games Strategies for Active Aircraft Protection from a Homing Missile,'' \textit{Journal of Guidance, Control, and Dynamics}, vol. 34, no. 3, pp. 761-773, June 2011.

\bibitem{Shaferman2} V. Shaferman, T. Shima, ``Cooperative Optimal Guidance Laws for Imposing a Relative Intercept Angle,'' \textit{Journal of Guidance, Control, and Dynamics}, vol. 38, no. 8, pp. 1395-1408, August 2015.

%%% more recent

\bibitem{Li} D. Li, J.B. Cruz, ``Defending an Asset: A Linear Quadratic Game Approach,'' \textit{IEEE Transactions on Aerospace and Electronic Systems}, vol. 47, no. 2, pp. 1026-1044, April 2011.

\bibitem{Kirchner} M. Kirchner, R. Mar, G. Hewer, J. Darbon, S. Osher, Y.T. Chow, ``Time-Optimal Collaborative Guidance Using the Generalized Hopf Formula,'' \textit{IEEE Control Systems Letter}, vol. 2, no. 2, pp. 201-206, April 2018.
%%% Assignment problem
\bibitem{Kuhn} H.W. Kuhn, ``The Hungarian method for the assignment problem,'' \textit{Naval Research Logisitcs Quarterly}, vol. 2, pp. 83-97, 1955.
\bibitem{Pentico} D. Pentico,``Assignment problems: A golden anniversary survey,'' \textit{European Journal of Operational Research}, vol. 176, no. 2, pp. 774-793, January 2007.

%%% Missile allocation problem
\bibitem{Matlin} S. Matlin, ``A Review of the Literature on the Missile-Allocation Problem,'' \textit{Operations Research}, vol. 18, no. 2, pp. 334-373.
\bibitem{Karasakal} O. Karasakal, ``Air defense missile-target allocation models for a naval task group,'' \textit{Computers \& Operations Research}, vol. 35, no. 6, pp. 1759-1770, June 2008.

%%% Assignment problem with collaborating agents
\bibitem{Younas} I. Younas, F. Kamrani, C. Schulte, R. Ayani, ``Optimization of Task Assignment to Collaborating Agents,'' in \textit{Proceedings of the 2011 IEEE Symposium on Computational Intelligence in Scheduling}, Paris, France, April 2011.

%%% LQDG/LQR theory
\bibitem{Pachter} M. Pachter, K.D. Pham, ``Discrete-Time Linear-Quadratic Dynamic Games,'' \textit{Journal of Optimization Theory and Applications}, vol. 146, no. 1, pp. 151-179, July 2010.
\bibitem{Anderson} B. Anderson, J. Moore, ``Optimal Control: Linear Quadratic Methods'', Englewood Cliffs, NJ, USA: Prentice-Hall, 1990.
\bibitem{Ho} Y.C. Ho, A.E. Bryson, S. Baron, ``Differential Games and Optimal Pursuit-Evasion Strategies,'' \textit{IEEE Transactions on Automatic Control}, vol. 10, no. 4, pp. 385-389, October 1965.
\bibitem{Bryson} A.E. Bryson, ``Dynamic Optimization'', Menlo Park, CA, USA: Addison Wesley Longman, 1999.
\bibitem{Hespanhalinear} J.P. Hespanha, ``Linear Systems Theory'', Princeton, NJ, USA: Princeton University Press, 2009.

%%% missile model
\bibitem{Tan} W. Tan, A.K. Packard, G.J. Balas, ``Quasi-LPV Modeling and LPV Control of a Generic Missile,'' in \textit{Proceedings of the 2000 American Control Conference}, Chicago, Illinois, USA, June 2000.

%%% overall missile system deisgn
\bibitem{Zarchan} P. Zarchan, ``Tactical and Strategic Missile Guidance, 6th Ed.'', Reston, VA, USA: American Institute of Aeronautics and Astronautics, 2012.

%%% PN guidance law
\bibitem{Palumbo1} N.F. Palumbo, R.A. Blauwkamp, J.M. Lloyd, ``Basic principles of homing guidance,'' \textit{Johns Hopkins APL Tech. Digest}, vol. 29, no. 1, pp. 25-41, 2010.
\bibitem{Palumbo2} N.F. Palumbo, R.A. Blauwkamp, J.M. Lloyd, ``Modern homing missile guidance theory and techniques,'' \textit{Johns Hopkins APL Tech. Digest}, vol. 29, no. 1, pp. 42-59, 2010.

%%% autopilot
\bibitem{Mracek1} C.P., Mracek CP, D.B. Ridgely, ''Missile longitudinal autopilots: connections between optimal control and classical topologies,'' in \textit{Proceedings of the 2005 AIAA Guidance, Navigation, and Control Conference}, San Francisco, California, USA, August 2005.
\iffalse
\bibitem{Mracek2} C.P., Mracek CP, D.B. Ridgely, ''Missile longitudinal autopilots: comparison of multiple three loop topologies,'' in \textit{Proceedings of the 2005 AIAA Guidance, Navigation, and Control Conference}, San Francisco, California, USA, August 2005.
\fi


\end{thebibliography}
\end{document}